\documentclass{aa}  
\usepackage{graphicx}
\usepackage{txfonts}

\usepackage[normalem]{ulem}

\begin{document}

   \title{Extreme ultra-soft X-ray variability in an eROSITA observation of the narrow-line Seyfert 1 galaxy 1H~0707$-$495}


\titlerunning{Ultra-soft X-ray variability, strong relativistic reflection and changing partial covering fraction}
      
   \authorrunning{Th. Boller, et al.}


   \author{Th. Boller,\inst{1}\thanks{E-mail: bol@mpe.mpg.de}, T. Liu\inst{1},  P. Weber\inst{2}, R. Arcodia\inst{1}, T. Dauser\inst{2}, J. Wilms\inst{2}, K. Nandra\inst{1}, J. Buchner\inst{1},   A. Merloni\inst{1}, M.J. Freyberg\inst{1}, M. Krumpe\inst{3},  S. G. H. Waddell\inst{1}
          }

   \institute{Max-Planck-Institut f\"ur extraterrestrische Physik, Giessenbachstrasse 1, 85748 Garching, Germany
         \and
             Dr.\ Karl Remeis-Observatory and Erlangen Centre for Astroparticle Physics,
Friedrich-Alexander-Universit\"at Erlangen-N\"urnberg, Sternwartstr.~7, 96049 Bamberg, Germany
         \and
         Leibniz-Institut für Astrophysik Potsdam (AIP), An der  Sternwarte 16, 14482 Potsdam, Germany
             }

   \date{Received September 1, 2020; accepted November 5, 2020}

 
  \abstract{
   {The ultra-soft narrow-line Seyfert 1 galaxy 1H~0707$-$495 is a well-known and highly variable active galactic nucleus (AGN), with a complex, steep X-ray spectrum, and has been studied  extensively with XMM-Newton.} 
   {1H~0707$-$495 was observed with 
    the extended ROentgen Survey with an Imaging Telescope Array (eROSITA)
    aboard the Spectrum-Roentgen-Gamma (SRG) mission
   on October 11, 2019, for about 60,000 seconds as one of the first 
calibration and pointed verification phase
(CalPV)  observations.
}
   {The eROSITA light curves show  significant variability in the form of a flux decrease by a factor of 58 with a 1~$\rm \sigma$ error confidence interval between 31 and 235. This variability is primarily in the soft band, and is much less extreme in the hard band.
No strong ultraviolet variability has been detected in simultaneous XMM-Newton Optical Monitor observations. 
The UV emission is $\rm L_{UV} \approx 10^{44}\ erg\ s^{-1}$, close to the  Eddington limit.
1H~0707$-$495 entered the lowest hard flux state seen in 20\,years of XMM-Newton observations. In the eROSITA All-Sky Survey (eRASS) observations taken in April 2020, the X-ray light curve is still  more variable in the ultra-soft band, but with increased soft and hard band count rates more similar to previously observed flux states. }
   {A model including relativistic reflection and a variable partial covering absorber is able to fit the spectra and provides a possible explanation for the extreme light-curve  behaviour. }
   {The  absorber is probably ionised and therefore more transparent to soft X-rays. This leaks soft X-rays in varying amounts, leading to large-amplitude soft-X-ray variability.}
}

   \maketitle
%

\section{Introduction}

All previous and present X-ray missions show that many narrow-line Seyfert 1 galaxies (NLS1;
see \citet{1985Osterbrock} and \citet{1989Goodrich}) have remarkable X-ray properties compared to Seyfert 1 galaxies with broader Balmer lines. 
Narrow-line Seyfert
1 galaxies are generally characterised by steep soft-X-ray spectra with photon indices  of up to about 5 from simple power-law fits. Detailed spectral modelling shows that NLS1s often have very strong soft-X-ray excess components compared to their hard X-ray tails. A clear anti-correlation is found between the ROSAT spectral softness and the H$\beta$ full-width at half-maximum intensity (FWHM) in type~1 Seyfert galaxies \citep{1996Boller} and quasars \citep{1997Laor}. This is remarkable as the X-ray emission from most type~1  Seyfert  galaxies originates predominantly from within a few to a few tens of Schwarzschild radii of their black holes, while Seyfert optical permitted lines are formed in a separate and significantly larger region. It appears that the anti-correlation between H$\beta$ FWHM and ROSAT spectral softness is part of a more general set of relations which involve the \citet{1992Boroson} primary eigenvector, and it has been suggested that NLS1s may be a subset of type 1 Seyfert galaxies that are accreting at relatively high fractions of the Eddington rate \citep{2005Tanaka}. Furthermore, NLS1s often show sharp spectral cut-offs in the high-energy spectrum, an observation that is still a point of controversy 
(see \citealt{2013Miller} and \citealt{2013Risaliti}). 
These objects also show remarkably rapid, large-amplitude X-ray variability. One spectacular object, the radio-quiet, ultra-soft NLS1, IRAS 13224$-$3809, shows persistent giant-amplitude variability events by factors of 35--60 on timescales of just a few days, most likely due to strong relativistic effects \citep{1997Boller}. The ROSAT HRI light curve of IRAS 13224$-$3809 is non-linear in character, suggesting that the X-ray emission regions on the accretion disc interact non-linearly or are affected by non-linear flux amplification. Dramatic flux and spectral variability has also been seen in many other NLS1s ---as described by some of the early ROSAT and ASCA publications--- such as for example  
Zwicky 159.034 \citep{1995Brandt},
WPVS007 \citep{1995Grupe}, 
1H 0707–495 \citep{1997Hayashida}, 
RE J1237+264 \citep{1995Brandt},
PHL 1092 \citep{1996Forster}, 
Mrk 766 \citep{1996Leighly}, and 
Ark 564 \citep{1994Brandt}.

1H~0707$-$495 has been observed with XMM-Newton for over 20 years. 
In this paper we report the eROSITA discovery of an extreme ultra-soft X-ray spectral state. The light curve is dominated by changes in the ultra-soft band, with much less pronounced variability in the hard X-ray band and no significant ultraviolet variability. We describe our data analysis in
Sect.~\ref{sec:data-extraction}, discuss the  light curve of the source in
Sect.~\ref{sec:lightcurve}, and then perform flux-resolved spectroscopy
in Sect.~\ref{sec:spectralanalysis}, where we show that the
variability can be explained by a temporally variable, ionised
absorber (Sect.~\ref{sec:interpretation}).

\begin{figure} 
\centering
\includegraphics[width=\columnwidth]{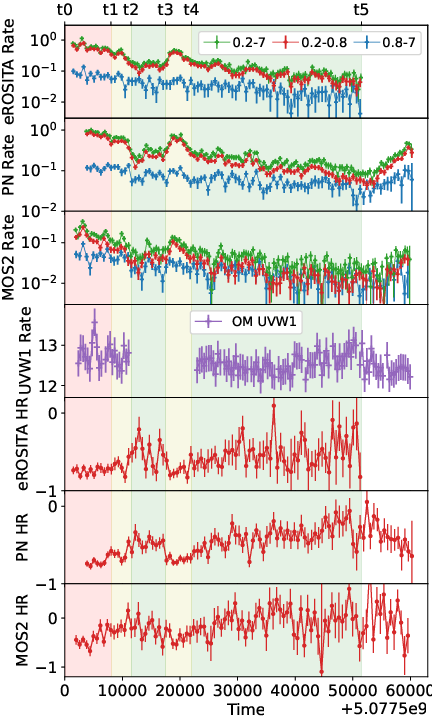}
    \caption{
    The top three panels display the background-subtracted light curves in the total (0.2--7.0\,keV), soft (0.2--0.8\,keV),  and hard (0.8--7.0\,keV) bands for eROSITA, EPIC-PN, and EPIC-MOS, respectively. 
    Large amplitude flux changes of about a factor $>50$ are detected in the total and soft X-ray light curves, with a normalised excess value of 34.8 and 44.6 $\rm \sigma$, respectively. The hard X-ray light curves and XMM OM light curve are much less variable, with normalised excess values of 2.1 and 1.7 $\rm \sigma$, respectively (c.f. Sect. 3.2 and 3.3). 
     The XMM OM light curve is shown in the fourth panel.
    The corresponding hardness ratios for the X-ray light curves are shown at the bottom.
    During the brightening the hardness ratio becomes softer and during low count rate intervals the hardness ratio is harder. 
    Three count rate states referred to as high, medium, and low are marked with light red, yellow, and green colours (Section~2.1).
    The X-ray total-band light curves have a bin size of 400s; the soft and hard bands have a bin size of 600s. 
}
    \label{fig:Fig1}
\end{figure}


\begin{figure} 
    \includegraphics[width=\columnwidth]{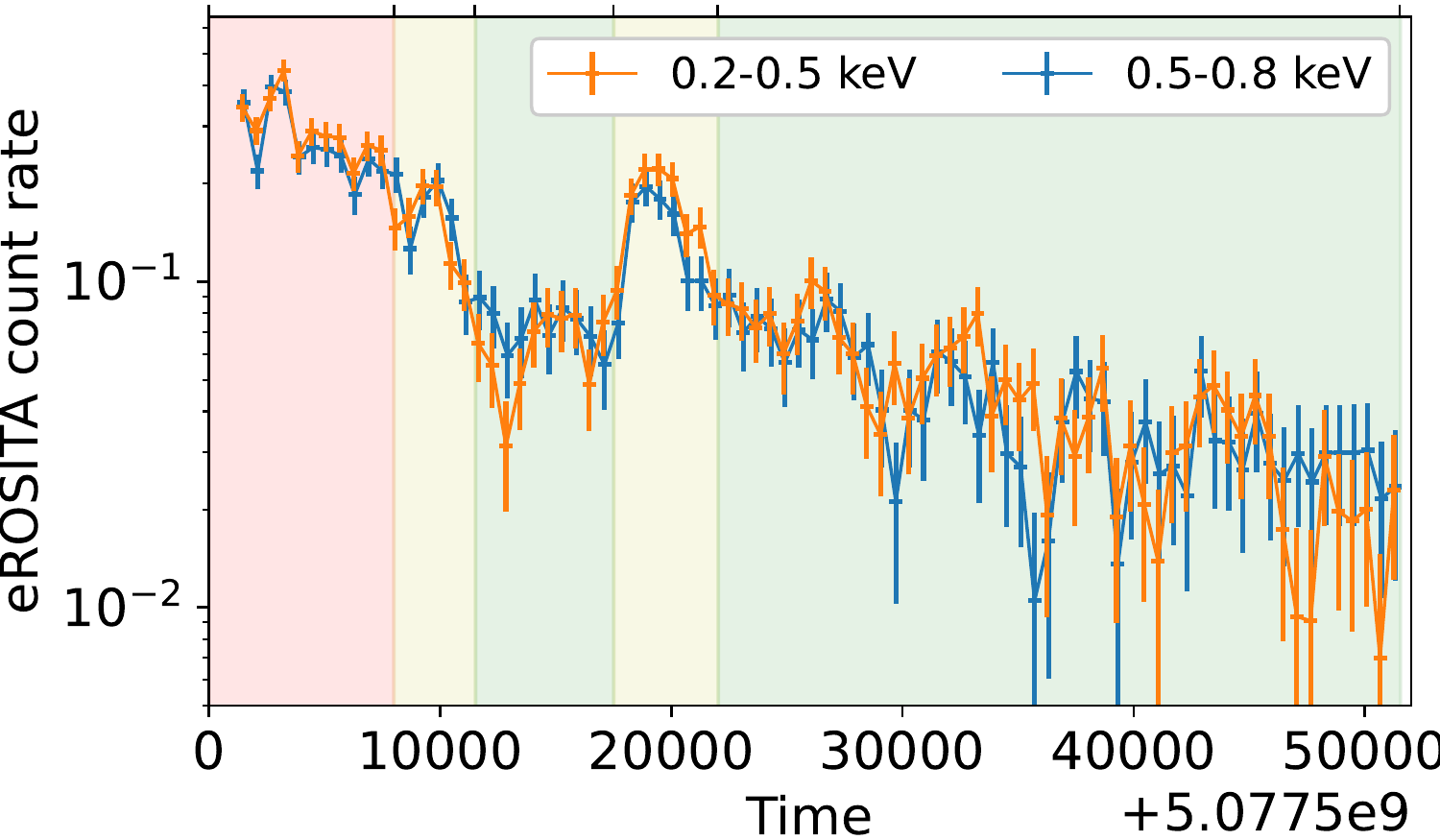}
    \caption{eROSITA light curves in the very-soft energy bands (0.2--0.5)\,keV (orange) and (0.5--0.8)\,keV (blue). The light curves are very similar.
}
    \label{fig:Fig2}
\end{figure}

\section{Data extraction}
\label{sec:data-extraction}

\begin{table}
\caption{%
Joint observations of SRG/eROSITA (ObsID 300003) and XMM-Newton (0853000101)
of 1H\,0707-495:
all eROSITA cameras were operated in FrameStore mode with the FILTER setup
(TM3 and TM4 were still switched off),
EPIC-pn and MOS2 cameras in LargeWindow mode with MEDIUM filter,
OM in Image/Fast mode in several exposures.
1H\,0707-495 was too weak for RGS1 and RGS2.
EPIC-MOS1 was still in its non-scientific CALCLOSED setup
due to a Single-Event-Upset in a previous observation.
}
\label{tab:ObsLog}
\begin{tabular}{llllr}
eROSITA  &  & \\ \hline\hline
            & Start$^a$ & End$^b$  &  expo$^c$\!\! & rate$^{d,e}$\,  \\
TM1         & 19:00:00  & 22:59:12 &  14.2         & 1.43$\pm0.11$ \\
TM2         & 20:06:50  & 22:59:12 &  10.3         & 1.20$\pm0.12$ \\
TM5         & 09:00:48  & 22:59:12 &  49.4         & 2.68$\pm0.08$ \\
TM6         & 09:00:48  & 22:59:12 &  50.3         & 2.91$\pm0.08$ \\
TM7         & 09:00:48  & 22:59:12 &  49.8         & 2.89$\pm0.08$ \\ \hline
XMM-Newton$\!\!\!$ & & \\\hline
EPIC-pn     & 09:38:14  & 01:21:40$^b$ &  52.0      & 14.70$\pm0.19$\\
EPIC-MOS2   & 09:06:53  & 01:14:49$^b$ &  57.3      & 3.49$\pm0.08$\\
OM          & 09:14:54  & 11:46:42     &            &                 \\
            & 14:58:42  & 01:21:10$^b$ &  44.0      & 12.56$\pm0.26$  \\
            \hline
\end{tabular}
\\

Notes:\\
$^a$ For eROSITA the start and end date is October 11, 2019 (times in UTC). \\
$^b$ For XMM-Newton the start date is October 11, 2019 and the end date is October 12, 2019 (times in UTC). Note the gap in OM exposure.\\ 
$^c$ Net exposure time is given in ks. \\
$^d$ For eROSITA and XMM-Newton the 0.5-0.8~keV net count rates are given in units of $\rm 10^{-2}\ counts\ s^{-1}$
after background subtraction. Note the different time ranges for the cameras.\\
$^e$ The OM count rates are not re-scaled. \\
\end{table}


\subsection{\textsl{eROSITA}}

eROSITA 
\citep{2012Predehl, 2012Merloni, Predehl2020}
is the primary instrument on the Russian  SRG mission (Sunyaev et al. 2020, in prep).
Following a 
CalPV phase
eROSITA is presently performing 4 years in scanning mode to create all-sky survey maps, superseding the ROSAT all-sky survey 
\citep{1984Truemper,1999Voges, 2016Boller}. 
Pointed observations are planned for after the survey phase.

1H~0707$-$495 was observed in a joint \textsl{XMM-Newton} and
\textsl{SRG} observation during the Performance
Verification (PV) phase of the \textsl{SRG} mission on October 11, 2019. As shown in Fig.~\ref{fig:Fig1} and Table~\ref{tab:ObsLog}, eROSITA started the observation slightly before \textsl{XMM-Newton} and also finished before. 
Telescope modules (TM)\,5, 6, and 7 were active during the entire
observation, whereas TM\,1 and TM\,2 were only activated for the last
16\,ks and 10\,ks of the observation, respectively, when the
source was essentially static. 1H~0707$-$495 was also observed during the first eROSITA all-sky survey (eRASS1) between April 26 and 29, 2020, for a total exposure of 407\,s. All cameras were active and the total number of counts is $\sim 400$.

Prepared event data were retrieved from the \textsl{C945} version of the standard processing for eROSITA products. \textsl{srctool} version 1.49 from the \textsl{eROSITA Science Analysis Software System} (eSASS) was used to extract light curves, spectra, and the necessary auxiliary files for data analysis 
 \citep{2018Brunner,2020Brunner}.
For the PV observation, data products were extracted with a source extraction circle of $60''$ in  radius and a background extraction annulus with inner and outer radii of $140''$
and $240''$, respectively, excluding nearby contaminating sources. For eRASS1 data, the background annulus radii were extended to $230''$  and $595''$ to increase the count statistics, albeit still excluding contaminating sources. 

As shown in Fig.~\ref{fig:Fig1}, for further characterisation we divide the PV observations into five count-rate-selected time intervals using six time points $t_0 \dots t_5$ . 
The [$t_0,t_1$] section is selected as the high-count-rate state,
the [$t_1,t_2$] and [$t_3, t_4$] sections are referred to as medium-count-rate states, 
the [$t_2, t_3$] and [$t_4, t_5$] sections are low-count-rate states,  
where 
$t_0=58767.361111,
 t_1=58767.453704,
 t_2=58767.494213,
 t_3=58767.563657,
 t_4=58767.615741, 
 t_5=58768.020833 (MJD)$.
 Flux-resolved spectral analysis is commonly used for highly variable objects, as applied for example by \cite{2019Kammoun} for NGC 4395. The time sequence applied for 1H~0707$-$495 in this paper is high-medium-low-medium-low.




\subsection{\textsl{XMM-Newton}}


The observation of 1H~0707$-$495 by \textsl{XMM-Newton} \citep{2001Jansen} started on 
October 11, 2019, and lasted for 60700 seconds until October 12, 2019 (ObsID 0853000101).
Extraction of the data was performed using the \emph{XMM Science Analysis System} (SAS) version 18.0.0.
For EPIC-pn \citep{2001Strueder}, the source and background photons were taken from circular
regions with radii of $35''$ and $106''$, respectively. The background
area was chosen on the same CCD chip as the source, and was chosen to be empty of other sources and exclude gaps in the CCD.
The same applies for data taken from MOS2
\citep{2001Turner},
where source and background
regions of $17''$ and $55''$ in radius were used.
MOS1 did not deliver any science products during the observation.
The source signal is too weak for an extraction from the Reflection Grating Spectrometer (RGS; \citealt{2001Herder}).

The \textsl{XMM-Newton} Optical Monitor \citep[OM;][]{Mason+2001:OM} covered the entire joint observation with 11 UVW1 exposures in Image mode, 10 of which were also taken in Fast mode. We processed the OM data using the tasks \texttt{omichain} and \texttt{omfchain} of SAS version 18.0.0. The standard and recommended procedure was adopted and the output products were checked following the list of known caveats and visual tests advised in the guides\footnote{\url{https://www.cosmos.esa.int/web/xmm-newton/sas-threads}}. We compared the surface brightness of 1H~0707$-$495 with two sources with high proper motion in the field, taken with the same aperture and with a similar count rate. The radial emission profiles were found to be very similar, and therefore the source can be considered close to point-like and with minor host contamination \citep[see also][]{Leighly+2004:1H0707_hubble}, also validating the automated coincidence loss correction in \texttt{omichain}.

\subsection{Analysis methods}
The \textsl{Interactive Spectral Interpretation System} (ISIS) in combination with \texttt{HEAsoft} version 6.27.2 was used for the analysis of spectra and light curves. 
\citet{Cash1979} statistics have been used for spectral fitting, as the signal to noise ratio (S/N) is insufficient for $\chi^2$ statistics.


\section{Light-curve analysis}
\label{sec:lightcurve}

In the following we describe the unique X-ray properties detected in the eROSITA observations. The results obtained from the simultaneous XMM-Newton observations are also discussed. We also report the first eROSITA all-sky survey observations performed in April 2020.

\subsection{Detection of large-amplitude flux changes}


During the eROSITA observations, 1H~0707$-$495 showed a dramatic flux
drop in about one day (see Fig.~\ref{fig:Fig1}).
The source is brightest at the beginning of the observations, with  rapid fluctuations in count rate, followed by a subsequent decline in count rate going down close to the background level. 
The highest count rate detected in the eROSITA (0.2--7.0)\,keV light curve is
$1.112\pm0.064\ \mathrm{counts}\ \mathrm{s}^{-1}$.
The corresponding lowest count rate is 
$0.019\pm0.014\ \mathrm{counts}\ \mathrm{s}^{-1}$. The resulting mean amplitude variability is a factor of 58, with a 1 $\rm \sigma$ error confidence interval with factors between 31 and 235.
Similar large-amplitude count-rate changes are deduced from the EPIC-pn and XMM-MOS2 light curves, where the lowest count rate values are consistent with the corresponding background values.


\subsection{Energy dependence of the variability}



















\begin{figure} 
    \includegraphics[width=\columnwidth]{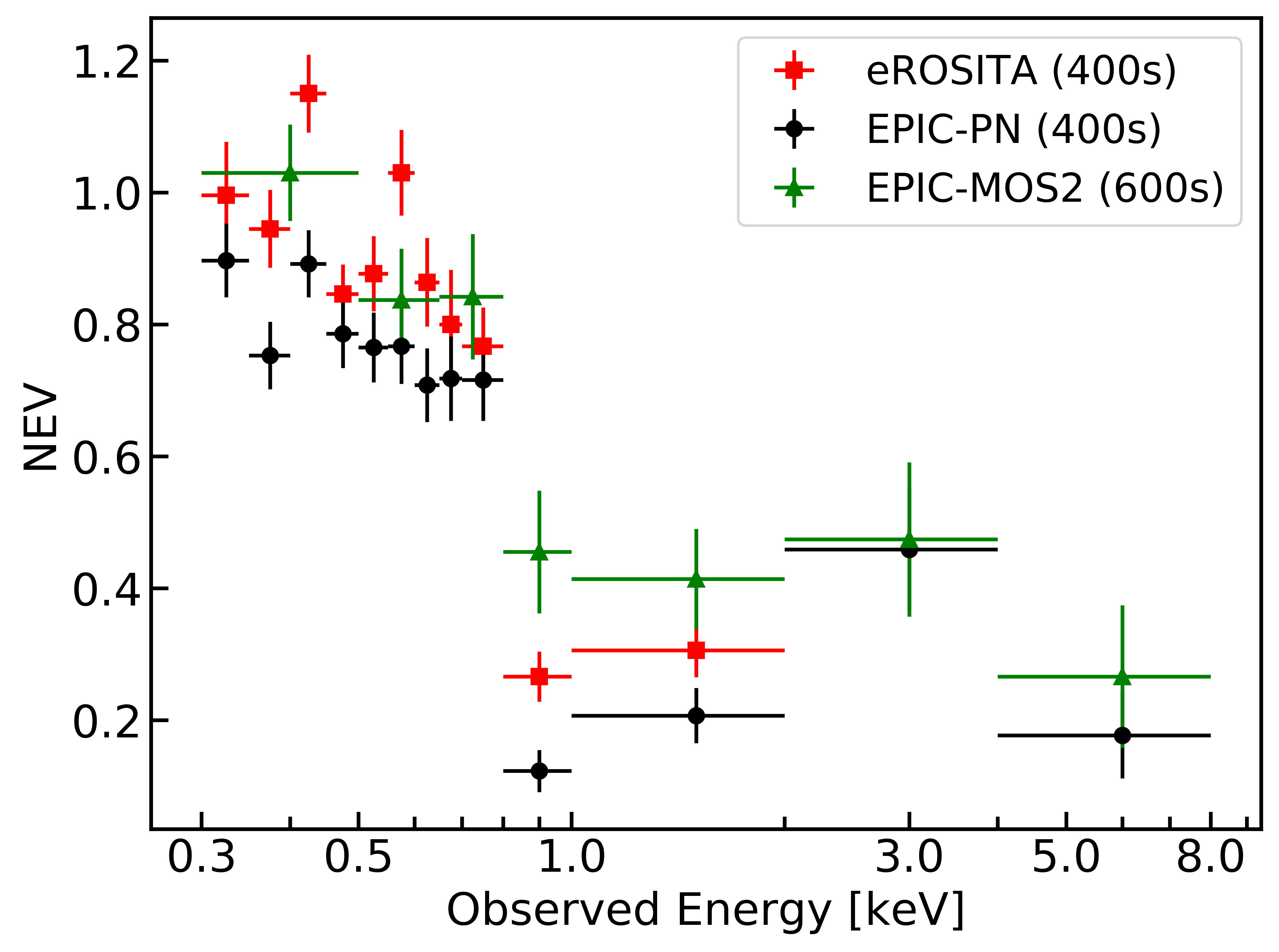}
    \caption{ 
    Normalised excess variance spectra calculated for eROSITA, EPIC-pn, and EPIC-MOS2. All spectra show the same shape, with significant variability up to 0.8\,keV followed by a sharp drop downwards between 0.8 and 2.0\,keV. }
    \label{fig:nev}
\end{figure}

Figure~\ref{fig:Fig1} presents the eROSITA and \textsl{XMM-Newton} light curves in the total (0.2-7.0 keV), soft (0.2--0.8\,keV), and hard (0.8--7.0\,keV) energy bands (c.f. Fig.~\ref{fig:nev} for motivation for these energy band selections). 
The soft variability appears similar to the total band variability, with count-rate changes of a factor greater than 50.
In the hard band, the variability amplitude is about a factor of 10, obtained from the XMM MOS2 light curve.
The normalised excess variance (NEV) 
is a powerful and commonly used method to test whether a time series is significantly variable above a certain threshold  \citep[e.g.][]{1997Nandra,2003Vaughn,2004Ponti}. The NEV values have been calculated for the  total, soft, and hard band eROSITA light curves based on Eqs. 1 and 2 of \citet{2016Boller}. Both the total and soft band eROSITA light curves are highly variable with NEV values of 34.8 and 44.6 $\rm \sigma$, respectively. The NEV value for the hard eROSITA light curve is $\rm 2.1 \sigma$, quantifying the higher amplitude variability in the  soft and total bands compared to the hard energy band.





We further analysed the soft X-ray light curve in the energy
bands (0.2--0.5)\,keV and (0.5--0.8)\,keV (see Fig.~\ref{fig:Fig2}, top
panel, for the eROSITA light curves). Interestingly, the light curve appears almost identical in both bands. 
Both are significantly variable with NEV values of 31.3 and 21.8 $\rm \sigma$ for the (0.2--0.5)\,keV and (0.5--0.8)\,keV, respectively.
Above 0.8\,keV, the variability abruptly declines up to the highest energies probed (c.f. Fig.~\ref{fig:FigA3}).

The NEV values
 are then computed in energy-resolved bins to create NEV spectra for each detector. The results are shown in Fig.~\ref{fig:nev}. 
Larger time and energy bins are required for MOS2 given the lower number of counts. 
Normalised excess variance 
values may differ slightly between instruments due to varying observation exposures, and bin sizes.
However, all NEV spectra reveal dramatic variability below 0.8\,keV, with a striking drop off between 0.8 and 2.0\,keV. Any NEV values above 2.0\,keV could not be computed with eROSITA due to high background and very low variability, but for EPIC-PN and EPIC-MOS2, the variability increases slightly from 2 to 4\,keV before dropping again from 4 to 8\,keV. 

The strong soft X-ray variability is extreme in relation to the weak hard X-ray variability and the lack of ultraviolet variability (see following section). 
Such extreme ultra-soft and large-amplitude flux variability in active galactic nuclei has not been
detected with other X-ray observations so far.
Extremely large-amplitude variability has been observed in the past in objects such as IRAS 13224$-$3809 \citep{1997Boller}, 
GSN 069 \citep{2019Miniutti} and RX J1301.9+2747 \citep{2020Giustini}.
What distinguishes the discovery here is the presence of such variability in the soft X-ray band, with simultaneous observations that show the absence of such variations in the hard X-rays (above 0.8\,keV). 

\subsection{Absence of strong UV variability in \textsl{XMM-Newton} OM observations}
\label{sec:OM}

The source remained quite constant in the UV based at an NEV value of $\rm 1.7 \sigma$.
This is indicated  by both photometry from imaging and count rates from the timing light curve. The \texttt{omichain} photometry indicates a count rate level of $\sim 12.56\,$cts\,s$^{-1}$ and an AB magnitude of $\sim 15.8$. 
No reddening correction was applied and the OM UWW1 data indicate the observed count rate.
The \texttt{omfchain} light curve is shown in Fig.~\ref{fig:Fig1} with a bin time of 400 seconds. The light curve does not show significant variability during the $\sim60\,$ks observation, in contrast to the highly variable soft-X-ray light curve. There is an indication of one low-amplitude  count rate increase in the first OM exposure, but not of the order seen in soft X-rays. It is well known that in NLS1s, the optical-UV emission varies less than that in X-rays \citep[e.g.][]{Ai2013:optnls1}, a phenomenon also seen in 1H~0707$-$495 with the two very deep 500ks XMM-Newton observations from 2008 and 2010 \citep{Robertson+2015:uv_x_corr,Done+2016:1h0707spin}. 
The OM data for these deep observations are remarkably constant. The largest difference is 10\%\ in the UVW1.
In particular, the OM rates were observed at $\approx 11$ and $\approx 14\,$cts\,s$^{-1}$ in those two epochs, respectively \citep{Robertson+2015:uv_x_corr}, similarly to our 2019 observations. Moreover, \citet{Robertson+2015:uv_x_corr} found no evidence of strong correlations between UV and X-rays on timescales of less than a week \citep[but see][]{Pawar+2017:1h0707_uv_x}.

The flux at the effective wavelength of the UVW1 filter (i.e. 2910 Angstrom) was computed from the \texttt{omichain} count rates using the correction factors listed in the \texttt{OM\_COLORTRANS\_0010.CCF} calibration file. 
The average UVW1 flux throughout the whole observation is $1.83\times10^{-11}\,\mathrm{erg}\,\mathrm{cm}^{-2}\,\mathrm{s}^{-1}$; it is marginally higher with $(1.86\pm 0.01) \times10^{-11}\,\mathrm{erg}\,\mathrm{cm}^{-2}\,\mathrm{s}^{-1}$ in the high-flux state. For the low-flux state, the UVW1 flux is consistent with the average flux with
$(1.82\pm0.01)\times10^{-11}\,\mathrm{erg}\,\mathrm{cm}^{-2}\,\mathrm{s}^{-1}$.
%

\begin{figure} 
        \includegraphics[width=\columnwidth]{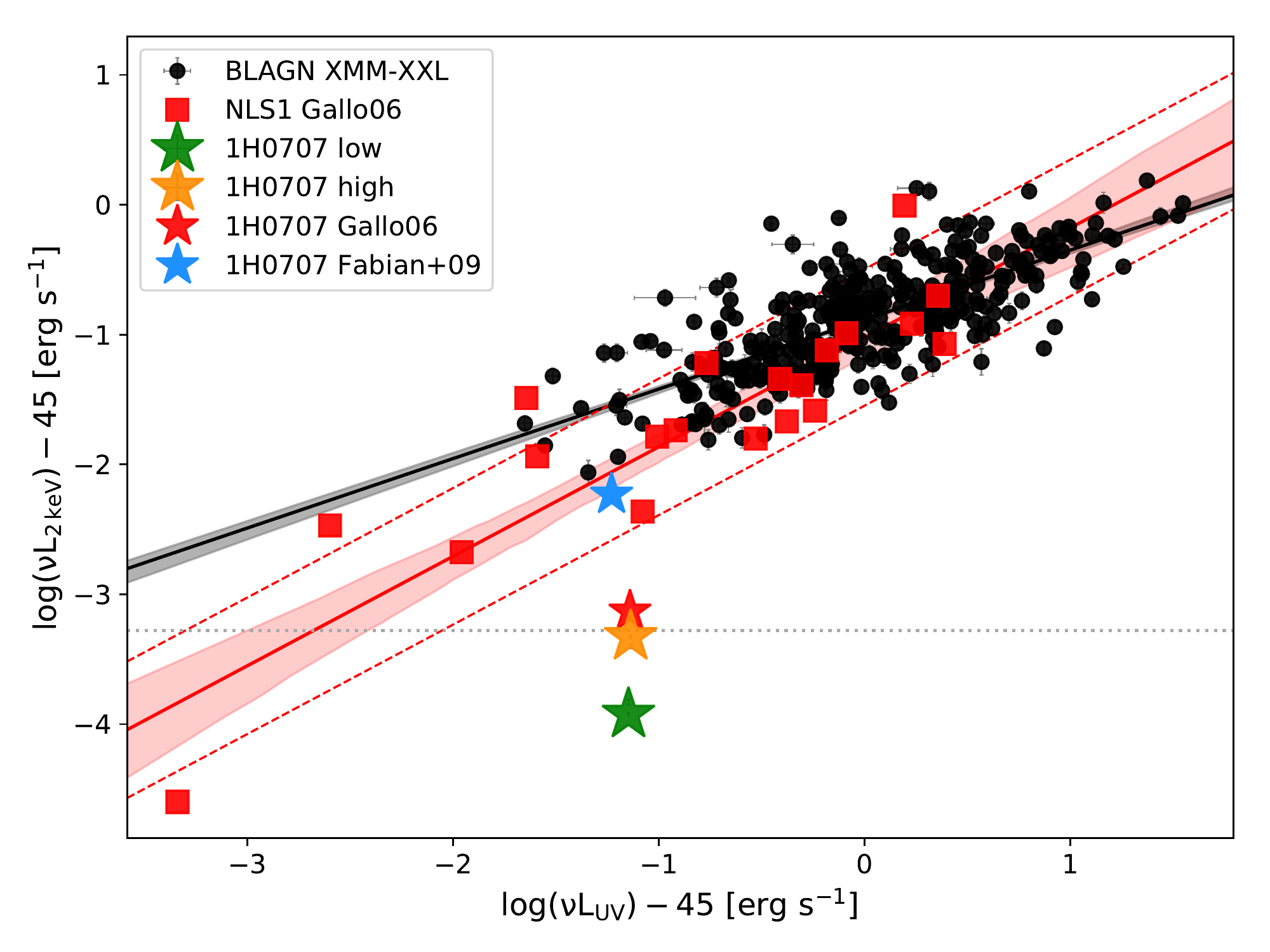}
    \caption{$L_X-L_{UV}$ plane with our 1H~0707$-$495 observations in the high-(orange star) and low-(green star)-count-rate state. 
    Results from an \texttt{emcee} \citep{Foreman-Mackey+2013:emcee} linear regression are shown for both broad-line AGNs 
    \citep[data points and regression as reported in][]{Arcodia+2019:lxluv},
    \citep[black circles;][]{Liu+2016:xxlL}
    and NLS1s (data points from \citealp{Gallo2006:nls1}, linear regression performed in this work). 
    We computed the X-ray luminosity at 2\,keV to compare with these literature measurements, while the UV proxy was computed at $3000$, $2500,$ and $2910$ Angstroms in XMM-XXL \citep{Liu+2016:xxlL}, NLS1 \citep{Gallo2006:nls1}, and our OM-UVW1 data, respectively. The best-fit linear regressions 
    \texttt{emcee} \citep{Foreman-Mackey+2013:emcee}
    are shown with a solid line, with corresponding 16th-84th percentile uncertainty intervals in a shaded area. The red dashed lines correspond to the fit intrinsic scatter of the NLS1 relation. The 2000 observation of 1H~0707$-$495 from \citet{Gallo2006:nls1} is shown with a red star instead of a red square; the median flux level from the 2008 observations reported in \citet{2009Fabian} is shown with a light blue star. The 2\,keV flux level during the eROSITA observation in the first all sky survey in April 2020 is shown with a grey dotted line.
    }
    \label{fig:lx_luv}
\end{figure}

In Fig.~\ref{fig:lx_luv} we show the OM-UVW1 and (EPIC-pn) 2\,keV luminosity for the high and low count rate state of 1H~0707$-$495 
compared to other NLS1s \citep[e.g.][]{Gallo2006:nls1} and to broad-line AGNs \citep[e.g.][]{Liu+2016:xxlL} in the $L_X-L_{UV}$ plane \citep[e.g.][and references therein]{Lusso+2016:lx_luv,Arcodia+2019:lxluv}.
%
%


From this comparison, it is clear that even the brighter state observed in our joint eROSITA/XMM-Newton observation is under-luminous in X-rays with respect to typical NLS1s and to past 1H~0707$-$495 observations as well \citep{Gallo2006:nls1,2009Fabian}. This indicates that we indeed observe an unusually X-ray-weak state of 1H~0707$-$495, especially when compared to other NLS1s given their UV emission. Remarkably, the UV level of 1H~0707$-$495 has remained within comparable values for the last $\sim20\,$ years \citep[e.g.][and references therein]{Robertson+2015:uv_x_corr,Done+2016:1h0707spin}.

\subsection{Comparison with 20 years of \textsl{XMM-Newton} observations and eROSITA all-sky survey observations}

\begin{figure}
        \includegraphics[width=\columnwidth]{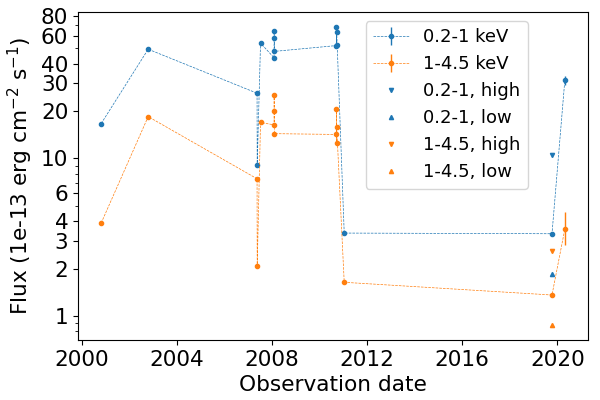}
    \caption{
    Soft and hard flux distribution of all \textsl{XMM-Newton} observations taken since the first observations in 2000 \citep{2002Boller} and as of 2019 and 2020 (this paper). The energy bands used are (0.2--1.0) and (1.0--4.5) keV, respectively. The  flux state of the hard band detected in 2019 simultaneous with the eROSITA observations is at an historical low. For the 2019 observations we also show the soft and hard high- and low-count-rate states.
    The data points for 2020 show the soft and hard fluxes obtained during the eROSITA all-sky survey observations.
    For illustration purposes we have connected the individual observations with the thin lines.
   }
    \label{fig:Fig4}
\end{figure}

The analysis and comparison with 20 years of \textsl{XMM-Newton} observations from 2000 \citep{2002Boller} and 2019 (this paper) reveals that 1H~0707$-$495 entered a historical low in hard flux band emission, first detected in simultaneous eROSITA  XMM-Newton observations (see Fig.~\ref{fig:Fig4}).
The lowest hard-band flux measured during the 2019 \textsl{XMM-Newton} observations is about $1.36\times10^{-13}\,\mathrm{erg}\ \mathrm{cm}^{-2}\ \mathrm{s}^{-1}$ (this paper),  about a factor of 15 lower 
than the highest hard flux value recorded, which was  about $2.0\times10^{-12}\,\mathrm{erg}\  \mathrm{cm}^{-2}\ \mathrm{s}^{-1}$  \citep{2012Dauser}. The soft band flux detected in the 2019 observations is $3.33\times10^{-13}\,\mathrm{erg}\ \mathrm{cm}^{-2}\ \mathrm{s}^{-1}$, consistent with the observations from 2010 \citep{2012Dauser}, but still also a factor of about 20 lower compared to the highest soft-band flux states of about  $7\times10^{-12}\,\mathrm{erg}\ \mathrm{cm}^{-2}\  \mathrm{s}^{-1}$.


\begin{figure}
        \includegraphics[width=\columnwidth]{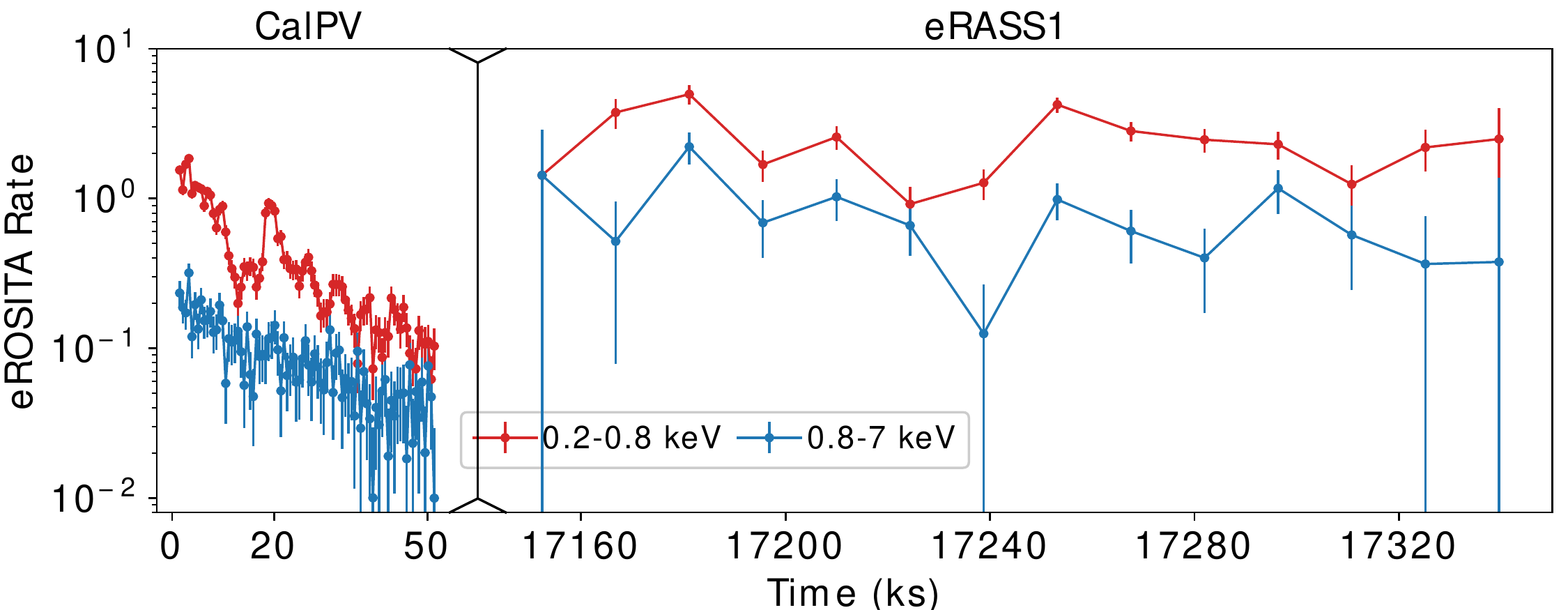}
            \caption{ eROSITA CalPV (observation on October 11, 2019) and eRASS1 (started on April 26, 2020) light curves in the 0.2-0.8 (red) and 0.8-7 (blue) keV bands. The CalPV light curves obtained using three cameras (TM5,6,7) are normalised to seven cameras, as all seven cameras were active during the eSASS1 observation. During the eROSITA all-sky survey observations both the soft- and hard-band count rates increased. The variability in the soft and hard bands is not as dramatic as that seen in the PV observations.
            }
    \label{fig:Fig5}
\end{figure}

1H~0707$-$495 was also observed during the first eROSITA all-sky survey scan (eRASS1) between April 26 and 29 in 2020. In each eRASS, every point in the sky is observed a few times (depending on the location of the source in the sky, the number increasing towards high ecliptic latitudes) for $\sim40$\,s every $\sim4$ hours. For 1H~0707$-$495, the net exposure is $407$\,s with 392 counts observed in total in the $0.2-7.0\,$keV band.
To convert counts to rates, we applied 
point spreak function (PSF)-loss
and vignetting corrections because the source enters the 
%
field of view (FoV) in each passage at different offset angles.
We extracted the light curves in the soft and hard energy bands from the survey data.  Figure~\ref{fig:Fig5}  shows the comparison between the eROSITA PV and eROSITA eRASS1 observations for the soft (0.2--0.8 keV) and hard (0.8--7.0 keV) energy bands on a 
logarithmic scale.
The soft light-curve count rate increased again during the eRASS1 observations, with less amplitude variability than seen in  the PV observations, and the hard band count rate also increased. We also report the related soft and hard band fluxes in Fig.~\ref{fig:Fig4}.

\section{Spectral analysis}
\label{sec:spectralanalysis}



\begin{figure}
        \includegraphics[width=\columnwidth]{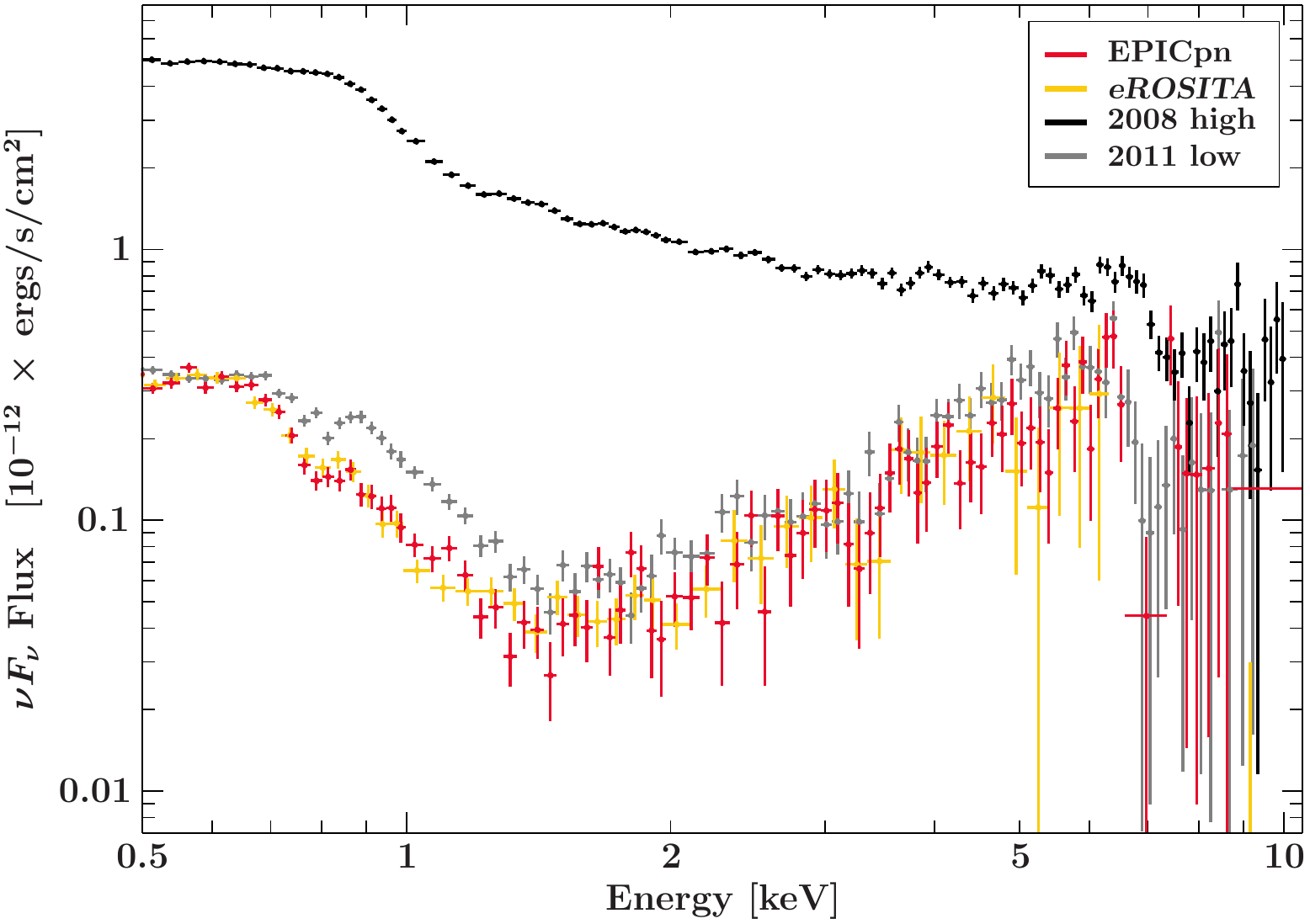}
    \caption{Comparison of the eROSITA and \textsl{XMM-Newton} observations using selected previous data; including the highest flux state observed in 2008 \citep[327\,ks][]{2009Fabian} and the low flux state in 2011 \citep[80\,ks][]{Fabian2012}. }
    \label{fig:spec-compare}
\end{figure}

In the previous section we show that the soft-band light curve displays
extreme and significant X-ray variability while the hard-band light curve is less variable. Figure~\ref{fig:spec-compare} shows the full eROSITA and
\textsl{XMM-Newton} EPIC pn spectrum of the 2019 observation. For
comparison, previously observed spectra of the highest and lowest flux
states from 2008 \citep[327\,ks,][]{2009Fabian} and 2011
\citep[80\,ks,][]{Fabian2012}, respectively, are shown. As seen in the overall flux evolution of 1H~0707$-$495
(Fig.~\ref{fig:Fig4}), the source in 2019 was caught in a very low flux state, with a
flux even lower than that observed in 2011 \citep{Fabian2012}.

The spectral shape of 1H0707$-$495 is characterised by a strong soft
component, followed by an almost flat part, and then a strong drop at around
7\,keV first reported by \citet{2002Boller}. A number of alternative models have been discussed to explain
this shape; see \citet{Fabian2012} for a discussion. These models
generally explain the spectrum as a combination of relativistic
reflection \citep[e.g.][]{Hagino2016} together with a strong soft
excess, as well as superimposed absorption features caused by a strong
wind \citep{2012Dauser}. In the following we use these earlier studies
to guide our spectral analysis, concentrating on the cause of the
spectral variability seen here. 
We note that other models based on inhomogeneous accretion flows (e.g. \citep{2006Merloni} have also been proposed to explain the complex spectral and timing properties of NLS1s with near-Eddington accretion flows.

\subsection{Relativistic reflection model}
\label{sec:relativistic_simple}

Due to the spectral similarity to the 2011 observation (see
Fig.~\ref{fig:spec-compare}), guided by \citet{Fabian2012} and in
agreement with analyses of the higher flux states \citep[e.g.
][]{2009Fabian,Zoghbi2010,2012Dauser} we describe the combined
0.5--10\,keV data with a relativistic reflection model. For this
analysis all spectra were optimally binned according to
\citet{Kaastra2016} and modelled using the \citet{Cash1979} statistic.

\begin{figure}
    \centering
    \includegraphics[width=\columnwidth]{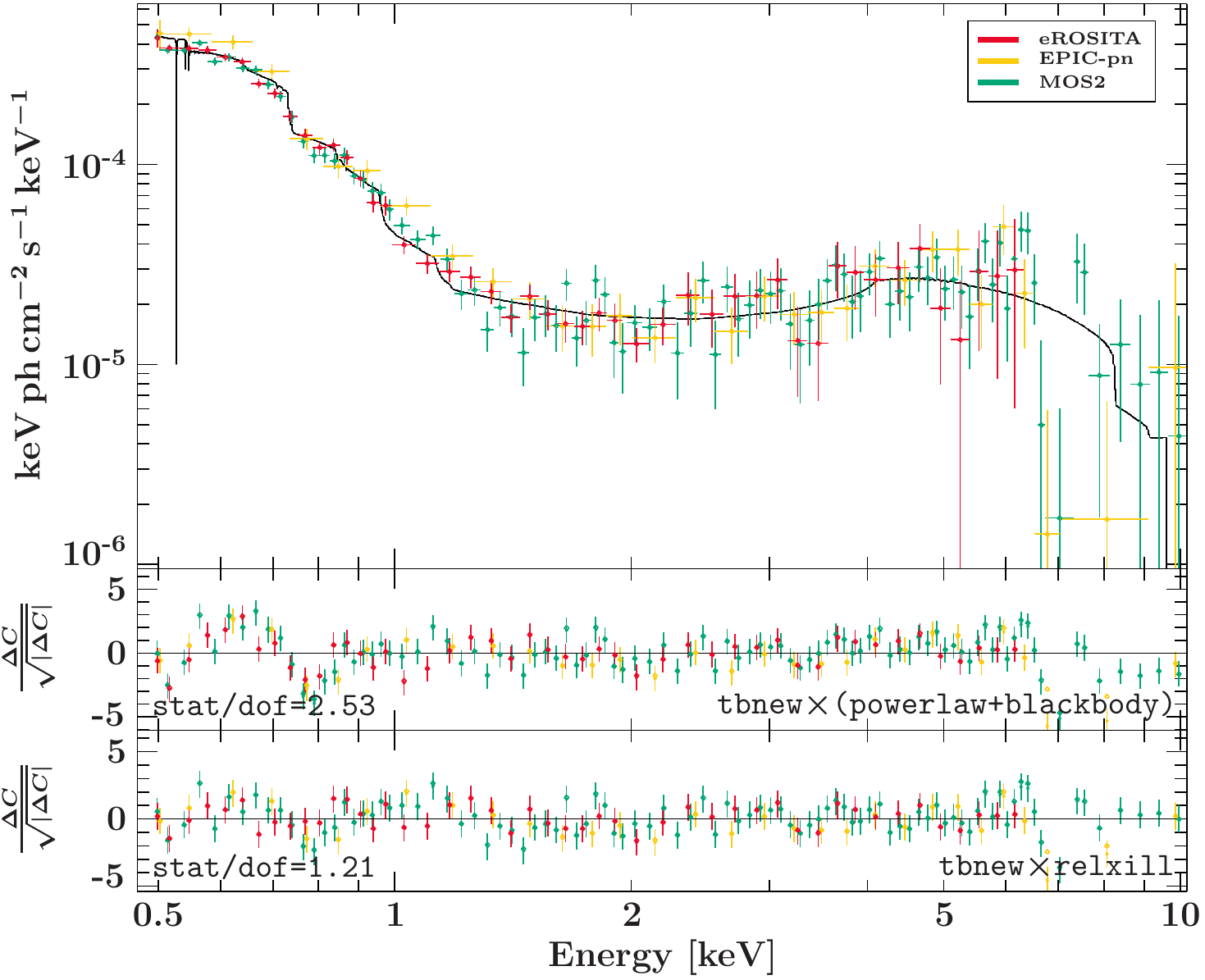}
    \caption{eROSITA, EPIC-pn and EPIC-MOS2 spectra of the entire
      observation, including the best fit relativistic reflection
      model (upper panel, brown, solid line). The middle panel shows the residuals
      of the best-fitting empirical model with a \texttt{blackbody}
      used to describe the soft excess and the lower panel shows the
      residuals for the relativistic reflection model. The spectra have been re-binned for plotting purposes only, for visual clarity. The spectra of EPIC-MOS2 and eROSITA were re-scaled
      to the flux normalisation of the EPIC-pn, using the best fit detector constants.
      }
    \label{fig:fit_all_data}
\end{figure}

Foreground absorption is accounted for using \texttt{tbnew}
\citep{Wilms2000} with abundances from \citet{Wilms2000},
fixing the equivalent hydrogen column density to the Galactic 21\,cm
equivalent width of $4.02 \times 10^{20} \mathrm{cm}^{-2}$
\citep{hi4pi}. The redshift to the source is set to $z=0.04057$
\citep{2009Jones}.
In order to account for potential differences in gain of the data due
to cross-calibration between instruments, multiplicative constants
(detector constants) for eROSITA ($C_\mathrm{eROSITA}$) and MOS2 ($C_\mathrm{MOS2}$)
with respect to pn were introduced into the models.
The relativistic reflection is described with the \texttt{relxill}
model \citep{Dauser2010,Dauser2013,Garcia2014}, which calculates the
relativistically smeared spectrum reflected from the innermost regions
of an ionised accretion disc. \texttt{relxill} is based on the
\texttt{xillver} model \citep{Garcia2013} for non-relativistic
reflection. Based on previous results, which suggest a very compact primary source
of radiation \citep{2012Dauser,Fabian2012} we use the \texttt{relxill}
lamp-post flavor, \texttt{relxilllp}, which assumes that the primary
source of the X-ray radiation is compact and located above the black hole on its rotational axis. The incident radiation from this source,
the so-called `corona', takes the form of a power law
with an exponential cutoff fixed at $E_\mathrm{cut}=300$\,keV. The
strength of the reflection component is parameterised by the
source intrinsic reflection fraction, $f_{\mathrm{refl}}$. It is defined in the frame of the primary source as the fraction of photons emitted towards the disc compared to the fraction emitted towards the observer \citep[see,][for a detailed definition]{Dauser2016}. As detailed in \citet{Dauser2014}, in the case of a low source height the strong light-bending effects would lead to most photons being focused on the disc and therefore easily to a reflection fraction of ten and larger.

Applying this relativistic reflection model to the eROSITA, EPIC-pn, and EPIC-MOS spectra  provides a good description of the data (with statistic/dof = 1.21). 
The spectra and the corresponding model are shown in 
Fig.~\ref{fig:fit_all_data} 
and the best-fit parameters are listed in Table~\ref{tab:tab_all_data}. We emphasise that no additional empirical black-body component is necessary to achieve a good fit when applying this relativistic reflection model to the data. Adding an additional low temperature ($kT \sim 0.1 \mathrm{keV}$) black-body component does not improve the fit statistics. For completeness, the comparison to a simple power law plus black body model is also shown in  Fig.~\ref{fig:fit_all_data}.

The best-fit parameters of this model are in good agreement with
previous results on relativistic reflection modelling of
1H~0707$-$495. Similarly to previous studies
\citep{Fabian2012,2009Fabian,2004Fabian,Kara2015}, iron is highly
overabundant, with an abundance of
$A_{\mathrm{Fe}}=10.0^{+0.0}_{-1.5}$ which is consistent with the upper
limit allowed by the reflection model. With
$\Gamma = 2.64^{+0.04}_{-0.08}$, the recovered photon index of the
incident power law is also in agreement with these earlier studies. The
spin parameter is well constrained, with a value of
$a=0.9960^{+0.0013}_{-0.0030}$, which is close to maximal spin, while the height
of the primary source, $1.39^{+0.023}_{-0.142}\,r_\mathrm{g}$, implies a
very compact X-ray source that is extremely close to the black hole.
These values are also consistent with earlier studies employing the
lamp-post geometry \citep{Fabian2012,2012Dauser,Kara2015}. While these
parameters tend to be consistent between the different earlier
observations, the inclination of the accretion disc was found to vary
widely, ranging from $23^{\circ}$ \citep{2002Fabian} up to
$78^{\circ}$ \citep{2012Dauser}. The value found in the present
analysis, $\theta = 73\fdg1^{+1.8}_{-1.6}$, is at the upper end
of this range. However, we emphasise that the self-similarity of
reflection spectra in the lamp-post geometry results in a degeneracy
between inclination and lamp-post height $h$ \citep{2012Dauser}, which
might be the reason for the large spread of observed inclinations. Recently, \citet{Szanecki2020a} applied their newly developed relativistic reflection model to an extended lamp-post source and confirm the compact nature of the corona  in agreement with the interpretation presented in the present study.

Interestingly, with $\log\xi = 0.73^{+0.12}_{-0.15}$, the ionisation
parameter of the accretion disc is low compared to previous
analyses \citep{2002Fabian,2004Fabian,Fabian2012,Hagino2016,Kosec2018}. Only
\citet{Kara2015} report a lower value of $0.3_{-0.2}^{+0.3}$ when also
using the \texttt{relxilllp} model on \textsl{NuSTAR} data from 2014. 

The reflection fraction is determined to be very high with $f_\mathrm{refl} = 46^{+13}_{-10}$, implying that most of the radiation emitted from the primary source is reflected on the disc and only a minor fraction is directly observed. This result is in agreement with previous observations starting with \citet{2002Fabian}, all consistently finding that 1H~0707$-$495 is extremely reflection dominated\footnote{in case a relativistic reflection model is used to describe the data; see above for alternative explanations.} \citep[see, e.g.][]{Kara2015}. Calculating the expected reflection fraction for such a point-like lamp-post source close to a very rapidly rotating black hole leads to values of $f_\mathrm{refl}^\mathrm{LP}=$12-20 \citep[see][]{Dauser2014}. This is in rough agreement with the high values we find, but still suggests a certain difference between the primary source in 1H~0707$-$495 and the standard lamp-post source.

Inspecting the residuals of the relativistic model in
Fig.~\ref{fig:fit_all_data} in more detail reveals that the drop in
flux around 6 to 7\,keV is not entirely correctly modelled. However, we note that a fast absorption by an ionised outflow as discovered
by 
 \citet{Done+07,2012Dauser} 
might explain why the model over-predicts the flux
around 7\,keV. Tailoring a disc wind model to the parameters of the 1H~0707$-$495
system, \citet{Hagino2016} were able to partly explain this drop as
ionised absorption seen under different velocities because of a wind cone
emitted between $45^\circ$ and $56^\circ$ which is intercepting the
line of sight. A\ detailed analysis of all available data by
\citet{Kosec2018} appears to be in support of the existence of an ultra-fast
stratified outflow in 1H~0707$-$495.

\begin{table}
\caption{Best-fit parameters and confidence intervals for the relativistic reflection model applied to the full observations, including both eROSITA and \textsl{XMM-Newton} data. We note that this is not our final model.}
\label{tab:tab_all_data}
\def\arraystretch{1.2}
\begin{tabular}{lc}
Parameter & Value and Confidence \\
\hline
\hline
spin$^a$    & $0.9960^{+0.0013}_{-0.0030}$ \\ 
$\theta$ [$^\circ$]  & $73.1^{+1.8}_{-1.6}$ \\ 
$\Gamma$        & $2.64^{+0.04}_{-0.08}$ \\ 
$h$ [$r_\mathrm{g}$]   & $1.385^{+0.023}_{-0.142}$ \\ 
$\mathrm{norm}_{\mathrm{relxilllp}}^b$     & $\left(3.9^{+26.0}_{-2.4}\right)\times10^{-4}$ \\ 
$f_\mathrm{refl}^c$       & $46^{+13}_{-10}$ \\ 
$A_\mathrm{Fe}^d$         & $10.0^{+0.0}_{-1.5}$ \\ 
$\log\xi^e$       & $0.73^{+0.12}_{-0.15}$ \\ 
$C_\mathrm{MOS2}$        & $1.14\pm0.07$ \\ 
$C_\mathrm{eROSITA}$     & $0.985^{+0.031}_{-0.030}$ \\
\hline
$C$-statistic/$\mathrm{dof}$ & $353.3/291$ \\
\hline
\end{tabular}
\par
Notes:\\
$^a$ dimensionless spin parameter $cJ/GM^2$ where $M$ is the black hole mass and $J$ is the angular momentum\\
$^b$ see \citet{Dauser2016}\\
$^c$ reflection fraction, see \citet{Dauser2014}\\
$^d$ iron abundance with respect to solar values \citep{Grevesse1996}\\
$^e$ ionisation parameter, defined as $\xi=4\pi F/n$ where $F$
is the incident flux and where $n$ is the particle density
\end{table}

\subsection{Spectra at high, medium, and low count rates}
In order to investigate the effect of the strong flux variability
during the observation,  we created three flux-resolved spectra, selected based on count-rate segments highlighted in Fig.~\ref{fig:Fig1}. The specific times of the selection are given in Sect.~\ref{sec:data-extraction}. As already seen by the detailed analysis of the light curves in different bands (see Sect.~\ref{sec:lightcurve}), the majority of the flux variability is detected below 1\,keV.

\begin{figure}
    \centering
    \includegraphics[width=\columnwidth]{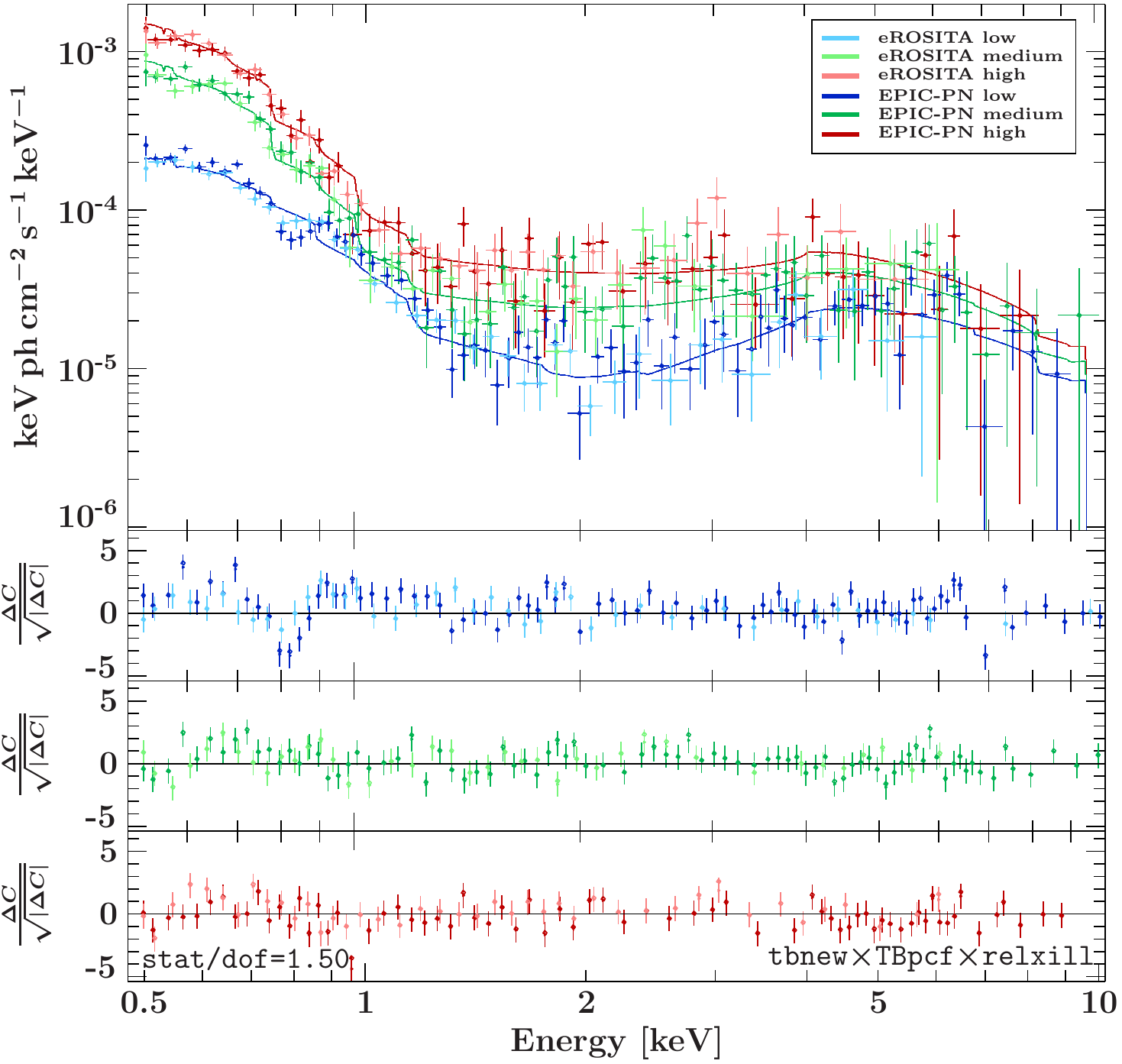}
    \caption{Count-rate-selected eROSITA and EPIC-pn spectra extracted from time windows highlighted in Fig.~\ref{fig:Fig1}. The underlying model is the best-fitting reflection model absorbed by a partial coverer with varying covering fraction. The corresponding parameters are listed in Table~\ref{tab:fit_states_pcf}. The data are strongly re-binned to facilitate visual inspection. Data of MOS2 are used in the spectral fits, but omitted in this plot to provide a clearer view. The lower panels show the residuals belonging to each flux state. The spectra of eROSITA are scaled according to the fitted detector constant to match the EPIC-pn data.
    The wind is directly detected from the more prominent edge at ~0.8 keV in the low-flux state.
    }
   \label{fig:ero_pn_comparison}
\end{figure}
Figure~\ref{fig:ero_pn_comparison} compares the eROSITA and EPIC-pn spectra in the three selected count-rate intervals. In the following sections, we explore the time evolution of the spectra with a partial covering model with relativistic reflection.

\begin{table}
\caption{Best-fit parameters and confidence intervals for our best-fit model fitting the data of the three flux states  simultaneously. The model consists of relativistic reflection in combination with a changing partial coverer. If only one value is given per row, this means it was tied between all spectra. See Table~\ref{tab:tab_all_data} for an explanation of the symbols used.}
\label{tab:fit_states_pcf}
\def\arraystretch{1.2}
\begin{tabular}{lccc}
Parameter & Low (1) & Medium (2) & High (3) \\
\hline
\hline
\multicolumn{4}{l}{Time-dependent parameters}\\
\hline
norm$_{\mathrm{relxilllp}}$ $\left(\times10^{-3}\right)$        & $3.5^{+1.6}_{-1.2}$ & 
$4.7^{+2.9}_{-2.1}$ & $5.6^{+2.9}_{-2.8}$ \\ 
$f_\mathrm{pc}$ & $0.75\pm0.06$ &$0.28^{+0.16}_{-0.22}$         & $ < 0.05$ \\ 
$\log\xi$     & $1.74^{+0.05}_{-0.04}$ & $0.64^{+0.17}_{-0.24}$ & $0.68^{+0.10}_{-0.21}$ \\ 
\hline
\hline
\multicolumn{4}{l}{Time-independent parameters} \\
\hline   
$N_{\mathrm{H}}$ $[10^{22} / \mathrm{cm}^2]$ & \multicolumn{3}{c}{$12^{+6}_{-4}$} \\ 
$A_{\mathrm{Fe}}$ & \multicolumn{3}{c}{$4.8^{+3.8}_{-1.2}$} \\ 
$\theta$ [deg] & \multicolumn{3}{c}{$74.4^{+1.4}_{-1.9}$} \\ 
spin & \multicolumn{3}{c}{$0.9968\pm0.0013$} \\ 
$\Gamma$& \multicolumn{3}{c}{$2.73^{+0.10}_{-0.09}$} \\ 
$h$ [$r_\mathrm{g}$] & \multicolumn{3}{c}{$1.25^{+0.11}_{-0.04}$} \\ 
$f_{\mathrm{refl}}$ & \multicolumn{3}{c}{$46^{+18}_{-16}$}\\ 
$C_\mathrm{MOS2}$ & \multicolumn{3}{c}{$1.04\pm0.06$} \\ 
$C_\mathrm{eROSITA}$ & \multicolumn{3}{c}{$0.954^{+0.030}_{-0.029}$} \\
\hline
stat/dof &  \multicolumn{3}{c}{$1209.3/806$} \\
\hline
\end{tabular}
\end{table}

\subsection{A changing partial coverer}

\begin{figure}
    \centering
    \includegraphics[width=\columnwidth]{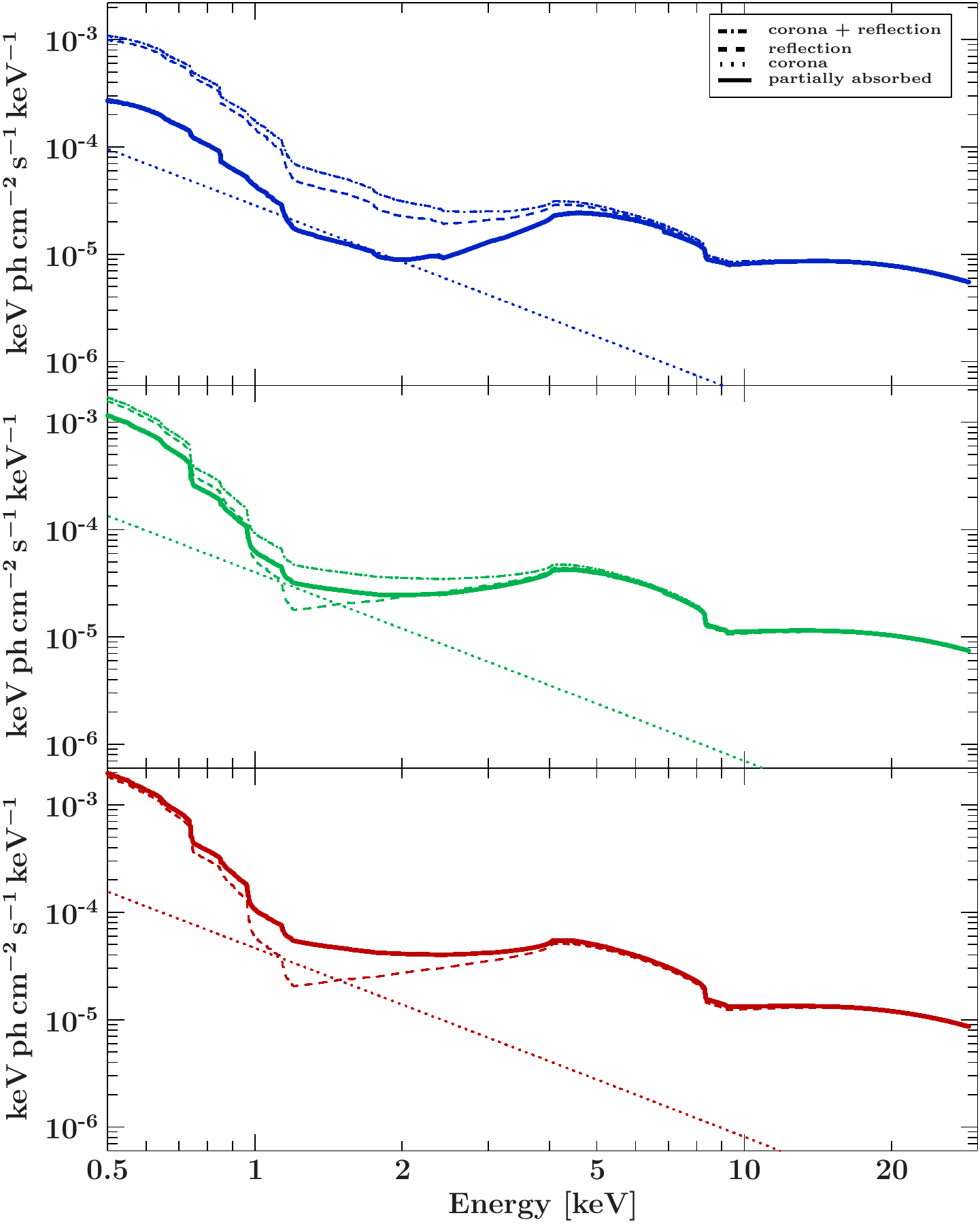}
    \caption{Decomposition of the model, for each count-rate state, into the radiation of the corona, which can be directly observed (dotted); the radiation reflected by the disc reaching the observer (dashed); the sum of both (dashed-dotted); and this sum partially absorbed (solid). The upper, middle, and bottom panels show the low, medium and high count-rate states, respectively. As the covering fraction in the high regime is zero, there is no absorption of the emitted primary and reflected spectra. The 
    Galactic foreground absorption is not shown in this plot. }
    \label{fig:states_models}
\end{figure}

Considering that X-ray absorption has a greater effect on the soft energies, we now check whether varying absorption can explain the
large changes observed in the soft flux of 1H~0707$-$495. To test
this hypothesis we employed the partial covering model \texttt{TBpcf}
\citep{Wilms2000} to act as a changing absorption component in the
line of sight towards the emission region. In order to test this
scenario, we fitted the spectra of all count-rate states simultaneously,
keeping all parameters of the continuum the same, including the column
density of the partial coverer, $N_{\mathrm{H}}$. The only parameters
that were allowed to vary between the two observations are the
ionisation parameter of the reflection model and the covering
fraction, $f_\mathrm{pc}$, of the partial coverer. All parameters as
determined from the best fit are listed in
Table~\ref{tab:fit_states_pcf}. A decomposition of the relativistic model for each flux state is shown in Fig.~\ref{fig:states_models}.

The best fit in this configuration yields $\mathrm{stat/dof} = 1209.3/806 = 1.50$, which means that the overall very strong variability can be explained
solely by a variation of the covering fraction of the absorber and by
a variation of the ionisation of the reflector. Over the course of the observation, the covering fraction 
$f_\mathrm{pc}<0.05$  for the highest flux states and then increases to
$f_\mathrm{pc}=0.28_{-0.22}^{+0.16}$ in the medium-flux spectrum 
and  $f_\mathrm{pc}=0.75 \pm 0.06$ in the low-flux spectrum.
We note that the time sequence of flux states is high-medium-low-medium-low. Because of low number statistics, we had to merge the two medium and the two low states in order to derive constrained covering fractions. 
Care is therefore needed when interpreting the covering fraction evolution.

For the parameters tied between the flux-selected spectra, a comparison with the best fit to the combined spectrum (see Table~\ref{tab:tab_all_data}) shows that
the photon index $\Gamma$, the reflection fraction
$f_{\mathrm{refl}}$, the inclination $\theta$, the lamp-post height
$h$, and the spin $a$ are consistent with the results from the
combined spectra. Only the iron abundance is now
reduced to a more reasonable value of $Z_{\mathrm{Fe}}=4.8_{-1.2}^{+3.8}$, possibly implying that the very high abundances found also in the earlier observations might be due to soft variability which was ignored in the analysis. 


Initial fits where we kept the ionisation parameter of the reflector  linked between flux-selected spectra did not yield a satisfactory description of the data, with significant residuals remaining in the 1\,keV band, which we attribute to emission in the Fe L band. We therefore allowed the ionisation parameter to vary between the three count-rate-selected spectra, which led to a good fit to the data. We note that we would not expect most other parameters such as the spin or the inclination to vary during the observation. The only parameter that has been suggested to change \citep[see, e.g.][]{Kara2015} is the height of the primary source. In our case a constant height satisfactorily describes the data and a potential additional change of the height cannot be detected. This can also partly be attributed to the lower S/N in these spectra.

The ionisation parameter changes from a consistent and fairly low ionisation for the high- and medium-flux-selected spectra ($\log\xi=0.68_{-0.21}^{+0.10}$ and $\log\xi=0.64_{-0.24}^{+0.17}$, respectively) to a larger value of $\log\xi=1.74_{-0.04}^{+0.05}$ in the low-flux spectrum.

As we discuss below, this change is likely not a physical change of the reflection component but the fit compensating for non-modelled ionisation in the absorber. 
In order to test a possible ionised partial coverer we used the model \texttt{zxipcf} 
\citep{Reeves08}. Unfortunately, because of the brevity of the observations, the data are insufficient for a signficant constraint to be put on the ionisation of the absorber. Even when freezing all \texttt{relxill} parameters and only allowing the normalisation, reflection fraction, and partial coverer fractions to vary freely, the ionisation in the medium flux state is completely unconstrained. In the low-flux state,
the ionisation parameter is weakly constrained to $\log\xi < 2$, but because of the low S/N, this value should be treated with care. Longer observations would be required to analyse the ionisation of the partial coverer. The low count-rate statistics do not allow us to constrain the ionisation of the absorber directly from the observations. 

\section{Physical Interpretations}\label{sec:interpretation}

\subsection{The extreme and varying UV-to-X-ray flux ratio}

One important new observational result is that within less than one day the ratio of UV to X-ray emission shows large variations. The UV emission is rather constant with $\rm L_{UV} \approx 10^{44}\ erg\ s^{-1}$, similar to the values reported by \citep{Done+2016:1h0707spin}, before applying bolometric corrections, which is close to the  Eddington limit.  On the other hand, the X-rays emission drops in amplitude by more than a factor of 50 (c.f. Fig.~\ref{fig:Fig1}).
A strongly varying X-ray flux during a constant UV flux was detected over timescales shorter than one day.
\cite{Buisson2017}  analysed a sample of 21 AGNs using data from the Swift satellite to study the variability properties of the population in the X-ray, UV, and optical bands. For 9 out of their 21 sources, 
the UV is lagging the X-rays.
For 1H~0707$-$495, the authors did not find strong correlations between the X-ray and the UV, similar to the results reported in this paper. \cite{Buisson2017} found 1H~0707$-$495 in a low-flux state during their Swift observations and argue that in such cases the source height of the illuminating corona is low, similar to the values reported in Table~\ref{tab:tab_all_data} in this paper, 
which makes it difficult to detect UV--X-ray time lags.
In the previous section, we infer that the X-ray variations are primarily due to varying covering fraction of a partial absorber. This does not seem to affect the UV, which implies that these are caused by independent physical processes. 
During the eROSITA CalPV observations, 1H0707$-$495 is extremely under-luminous in the X-ray compared to other NLS1s and BLS1s
(c.f. Fig.~\ref{fig:lx_luv}) as well as to the 1H~0707$-$495 high-flux-state
observations from \citet{2009Fabian}. 
This supports the interpretation that the X-rays are suppressed in this observation, and thus possibly absorbed, and therefore absorption-related changes could explain the variability.

\subsection{Changing partial covering fractions causing large-amplitude and ultra-soft count-rate variations}


The most important result of the analysis presented here is that the major source of variability observed in the spectrum can be explained by variation of the covering fraction of the  absorber.
Our spectral analysis shows that the variation of the X-ray spectrum is consistent with changes induced by a partial absorber of varying covering factor and constant column density in front of the X-ray-emitting corona and accretion disc. As expected, the covering fraction is increasing significantly with decreasing flux of the source. With $N_{\mathrm{H}} = 12_{-4}^{+6} \times 10^{22}\,\mathrm{cm}^{-2}$, the equivalent hydrogen column density of the partial coverer is consistent with that seen in typical AGN absorption events. \cite{2014Markowitz} find peak $N_{\mathrm{H}}$ column densities of 4--26 $\times 10^{22}\,\mathrm{cm}^{-2}$ in the largest sample of cloud obscuration events. Studying the long-term X-ray spectral variability of a sample of 20 Compton-thin type II galaxies, \cite{2020Laha} find 11 sources that require a partial-covering obscuring component in all or some of the observations. Not only are the $N_{\mathrm{H}}$ ranges quoted in both studies fully consistent with our derived value, but also the presence of a varying partial cover seems to be present in a significant fraction of AGNs.

 We note that there has been controversial discussion of whether a partial coverer in 1H0707$-$495 can explain the strong 7\,keV edge  \citep[e.g. ][]{2004Fabian, Gallo+2004, Done+07}. In our model, the partial coverer does not explain the 7\,keV edge. While this edge is mainly modelled by relativistically smeared reflection from the accretion disc, our partial covering model describes the varying absorption in the soft X-rays. In this paper we combine relativistic reflection very close to the black hole, that is, at a few $\rm R_G,$ with partial covering occurring at larger distances up to a few hundred $\rm R_G$. 
 From analyses of much longer observations \citep{2012Dauser,Kosec2018}, it is known that a strongly ionised wind is present in 1H0707$-$495. 
 The absorption feature around 0.8\,keV is evidence that this outflow is also present in the low flux state of our observation (c.f. Fig.~\ref{fig:ero_pn_comparison}). The wind is not detected in the higher flux states, as the outflowing winds are strongly flux dependent as shown for example by  \cite{2017Parker} and \cite{2018Reeves}.
As the existence of such an ultra-fast outflow (UFO) has been shown to be connected with the observed partial covering in other sources \citep[e.g. PDS~456, ][]{2018Reeves}, it is possible that the observed partial covering in the soft X-rays is connected to these previously detected UFOs. 
This absorption is likely connected to or even directly caused by the UFO detected previously.
The UFO will also affect the Fe-K region around 7\,keV \citep{Kosec2018}, but was not detected in our observations because of the lower S/N.

The change in partial covering fractions combined with UFO features may also explain the observed shape of the NEV spectra. On short timescales, the absorber is likely driving the variability, which is probably because of the small variations in ionisation and covering fraction as the material passes along the line of sight. As seen in Fig.~\ref{fig:states_models}, the absorber seems to affect the spectral shape between 0.3 and 4\,keV, which explains why these energy bins have higher NEV values. In particular, most of the variability is seen below 0.8keV, in agreement with what is seen in the light curve.

The NEV spectra also reveal very little variability in the 0.8-2.0\,keV and 4-8\,keV bands. This may be explained by the presence of UFO features in these energy bands. The outflow may be more stable on shorter timescales, instead varying on longer timescales. This behaviour would suppress the variability on short timescales in these energy ranges, explaining the drops in the NEV (c.f. Sect. 5.3 for a more detailed discussion on the connection between outflowing winds and partial covering).

However, at the same time, we also measure a change in the ionisation parameter of the relativistic reflection component. 
We consider it unlikely that this change of ionisation is indicative of
changes in the accretion disc, and is probably rather caused by the simplified (neutral) absorber model.
As discussed in Sect. 5.1, the data do not allow us to constrain the ionisation of the absorber. 

Due to lack of additional information, such as the ionisation of the absorber, it is not easy to estimate the distance and size of the obscuring cloud. Given the short timescale of the putative absorption event and the strong change of the covering fraction within 20-40\,ks, the absorber will probably be
much closer to the X-ray source than the BLR (see Sect. 5.3). This short distance makes it very likely that it will be partly ionised. 
However, ionised absorbers  are more transparent in the soft X-rays than neutral absorbers and therefore show leakage effects in the soft X-rays. The change in $\log\xi$ of the reflector seen here mainly affects the soft X-rays, and thus might mimic this effect of ionised absorption. We note that longer observations of such a partly obscured state would be necessary to constrain more detailed ionised absorption models for the partial coverer \footnote{We note that, e.g., \citet{Kosec2018} combined $>500\,$ks of data to detect the ionised outflow.}.

\begin{figure}
    \centering
    \includegraphics[width=\columnwidth]{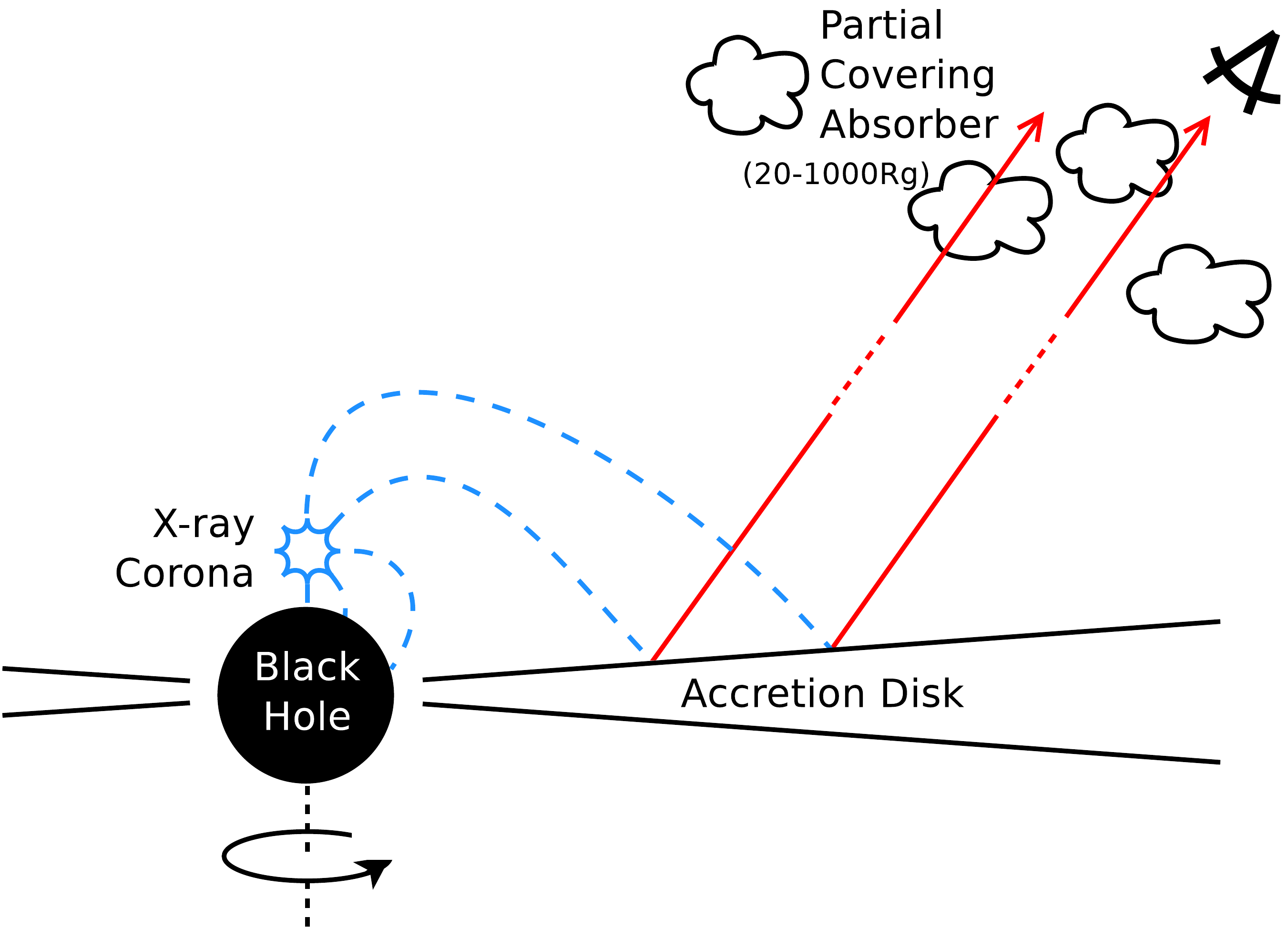}
    \caption{Illustration of the considered scenario. Above a spinning black hole, X-rays are emitted isotropically. Because of the compact corona very close to the black hole, the majority of the photons either hit the accretion disc or fall into the black hole. From the high- to the low-flux state, a partial coverer is obscuring and increasing part of the emitted X-ray radiation. 
    }
   \label{fig:illustration}
\end{figure}

An illustration of the changing partial coverer scenario with relativistic reflection is shown in Fig.~\ref{fig:illustration}. 
Because of gravitational light bending, the majority of the photons emitted from the corona are bent towards the black hole and onto the accretion disc  in  approximately equal parts \citep[c.f. Fig. 1 and 2 of][]{2014Dauser}). While in the high-flux state we have an unobscured view onto the inner parts of the accretion disc, partially covering clouds absorb the reflected spectrum in the lower flux states with increasing covering fraction for a decreasing observed soft X-ray flux.

\subsection{Partially covering absorbers and ultra-fast outflows}

Outflowing winds launched from the accretion 
disc
by radiation pressure or magnetic 
fields 
are considered as an important AGN feedback process. For radiation-pressure-dominated winds, outflows can reach velocities up to about 0.3 c and can drive substantial amounts of material into the interstellar medium. These winds have been discovered mainly based on XMM-Newton observations 
(e.g. \citealt{2003Poundsa, 2003Poundsb, 2003Poundsc, 2003King, 2003Reeves}).
Outflowing winds with such high velocities have been named as UFOs by \cite{2010Tombesi} in a systematic study of bright XMM-Newton AGNs.

Multiple outflow absorption lines have been detected in one of the most variable AGNs, IRAS 13224$-$3809, by \citep{2017Parker}. These latter authors argue that the X-ray emission from within a few gravitational radii of the black hole is ionising the disc winds up to hundreds of $\rm R_G$. It was also shown that the outflow absorption lines are strongly flux dependent, with strongest being found in the low-flux state and weakest in the high-flux state, which is due to increasing ionisation towards higher flux values. When the ionisation becomes sufficiently high, the outflow may become `over-ionised' and may no longer be visible. Such a scenario was also discussed by \cite{2019Gallo}, where absorption features were detected in the beginning of a flare in the NLS1 Mrk 335 but not in the brightest prolonged flare states. 

Ionised outflowing winds have been connected to absorbing partial covering by for example \cite{2018Reeves} and references therein. These latter authors argue that the outflowing wind is inhomogeneous and more complex than a simple homogeneous outflow, which is capable of partially covering the X-ray source. In this scenario, the X-ray absorption depends on the ionisation state, the distance of the absorber, and the covering fraction. 

The two XMM-Newton observations of PDS 456 reported by \cite{2018Reeves} were made over two consecutive satellite orbits. These latter authors argue that much of the spectral variability between the observations appears to be reproduced by the variability of the low-ionisation partial covering absorber, which is primarily driven by a change of the covering fraction. This appears consistent with the low-flux states and the varying covering fractions reported for PDS 456 and now for 1H~0707$-495$ in this paper.

Partial covering absorbers have been put into context with UFOs and winds in several other papers.
\cite{2020Reeves} provide a further exploration for the spectral shape and variability of PDS 456, noting in particular the significant differences in the soft-band fit when using neutral and ionised partial-covering components. There are many other works that analyse the soft- and hard-band emission and absorption features in PDS 456, concluding that an outflowing absorber can explain these features as well as the observed variability (e.g. \citealt{2016Matzeua, 2016Matzeub, 2018Parker}).

A larger sample of Seyfert galaxies analysed in \cite{2013Tombesi} also reveals that many AGNs that display UFO signatures also show evidence for warm absorption, and based on their observed properties, these latter authors propose that these may actually be part of a single large-scale outflow. Simultaneous observations of absorption and outflowing components are also presented for individual sources, including Mrk 335 (\citealt{2019Longinotti}, but see also \citealt{2019Gallo}) and PG 1211+143 (\citealt{2016Pounds}). This lends further support to the idea that such components may be physically linked and appear simultaneously, as in the observations presented in this work.

\subsection{Speculations on the partial coverer size and location}

Because the absorption is only partial, we can place limits on the projected size of the absorber. For such an extreme configuration of compact corona and large black hole spin, most of the observed flux is due to reflection from within a radius of 5-10$\;R_g$ around the black hole \citep{Dauser2013}, suggesting that the absorbing structure is smaller than this scale.


From Fig.~\ref{fig:Fig1} we estimate that a first obscuration event is seen between times $\rm t_0$ and $\rm t_3$ for about 20000 seconds where the count rate decreases from the highest count-rate state to the lowest count-rate state.  Between $\rm t_3$ and $\rm t_4$, the count rate increases again but probably with the covering fraction found in the medium-flux state. A third obscuration event might be detected from $\rm t_4$ until the end of the eROSITA observations, where the source is found in the lowest count-rate state with the highest covering fraction.

To estimate the distance of the absorbing cloud we adopt Eq.~2 of \citet{2017Beuchert}.
Considering cloud number densities $n_{\mathrm{H}}$ from $10^9\,\mathrm{cm}^{-3}$ to $10^{10}\,\mathrm{cm}^{-3}$ yields distances from $11\,R_{\mathrm{G}}$ to $1100\,R_{\mathrm{G}}$ for the first obscuration event, which assumes Keplerian orbits, corresponding to an orbital velocity of 0.2\,c to 0.02\,c. To change the covering fraction from the $\rm t_0$ to  $\rm t_3$ from less than 0.1 to 0.73 within about 20000 seconds, the projected length of the absorber is in the range of $1.2 \times 10^{13}$\,cm to $1.2 \times 10^{14}$\,cm, or 0.03 to 0.3 light days.
This seems reasonable but we avoid further speculation on distances and sizes of the absorber in order not to over-interpret the available data.

\section{Summary}

We detected large-amplitude variability with changes of more than   a factor of 50 in the eROSITA light curves.
The soft band (0.2--0.8 keV) dominates the variability, while in the hard band (0.8--7.0 keV) the variability is much less extreme.
We further analysed the soft X-ray light curve, dividing the soft-band light curve into two very soft X-ray light curves in the energy bands (0.2--0.5) keV and (0.5--0.8) keV. 
Both ultra-soft X-ray light curves are close to identical in their count-rate distributions as a function of time. Above 0.8 keV, the variability abruptly declines up to the highest energies. This behaviour is further confirmed by analysing the normalised excess variance spectra (see Fig.~\ref{fig:nev}), where all instruments detect significant variability up to 0.8\,keV, followed by a sharp drop off. 
This is the first time that such large-amplitude ultra-soft variability has been detected  with eROSITA observations in AGNs. There are two sources with similar but less extreme ultra-soft  variability behaviour,  GSN 069 \citep{2019Miniutti} and RX J1301.9+2747 \citep{2020Giustini}.

No significant variability was detected in the UV in the XMM-Newton OM observations. 
The UV emission is relatively constant with $\rm L_{UV} \approx 10^{44}\ erg\ s^{-1}$, similar to the values reported by \citet{Done+2016:1h0707spin} which is close to the  Eddington limit. In the  combined eROSITA and \textsl{XMM-Newton} observation, 1H~0707$-$495 was caught in a historically low hard-flux state, similar to the low flux state reported by \citet{Fabian2012}. 

We use the relativistic reflection model 
\texttt{relxill} \citep{Dauser2010,Dauser2013,Garcia2014}
to fit the data, and find parameters in good agreement with these latter publications. 
Spectral changes were investigated by constructing three count-rate-selected spectra, to which the partial covering model \texttt{TBpcf}
\citep{Wilms2000} in combination with the \texttt{relxill} model was applied.
The majority of the change in spectral shape during this observation can be fully explained by a varying covering fraction, rather than varying column density or ionisation.  We conclude that the large variability of the soft flux detected in the light curve is fully consistent with the varying covering fraction interpretation. These findings strongly suggest that the variability and the large change in soft flux during the observation is caused by an AGN obscuration event. Further evidence in support of this scenario is added by the fact that when 1H~0707$-$495 was observed 6~months later during the eRASS1 survey, its flux returned to the flux level above our unobscured model. \textsl{eROSITA} will observe 1H~0707$-$495 another seven times every 6 months until completing its all-sky scanning mission.

\begin{acknowledgements}
We are grateful to M. Page for his help in the \textsl{XMM-Newton} OM data analysis. We thank the \textsl{XMM-Newton} PS N. Schartel for accepting the simultaneous \textsl{XMM-Newton} observations. MK acknowledges support by DFG grant KR 3338/4-1.

We thank the anonymous referee for their careful reading of the submitted manuscript, and for their very helpful comments and suggestions.

This work is based on data from eROSITA, the primary instrument aboard SRG, a joint Russian-German science mission supported by the Russian Space Agency (Roskosmos), in the interests of the Russian Academy of Sciences represented by its Space Research Institute (IKI), and the Deutsches Zentrum für Luft- und Raumfahrt (DLR). The SRG spacecraft was built by Lavochkin Association (NPOL) and its subcontractors, and is operated by NPOL with support from the Max-Planck Institute for Extraterrestrial Physics (MPE).

The development and construction of the eROSITA X-ray instrument was led by MPE, with contributions from the Dr. Karl Remeis Observatory Bamberg, the University of Hamburg Observatory, the Leibniz Institute for Astrophysics Potsdam (AIP), and the Institute for Astronomy and Astrophysics of the University of Tübingen, with the support of DLR and the Max Planck Society. The Argelander Institute for Astronomy of the University of Bonn and the Ludwig Maximilians Universität Munich also participated in the science preparation for eROSITA. 

The eROSITA data shown here were processed using the eSASS/NRTA software system developed by the German eROSITA consortium.

\end{acknowledgements}

\bibliographystyle{aa}
\bibliography{39316corr.bib}

\appendix

\section{Amplitude variability as a function of energy}

\begin{figure} 
 \includegraphics[width=\columnwidth]{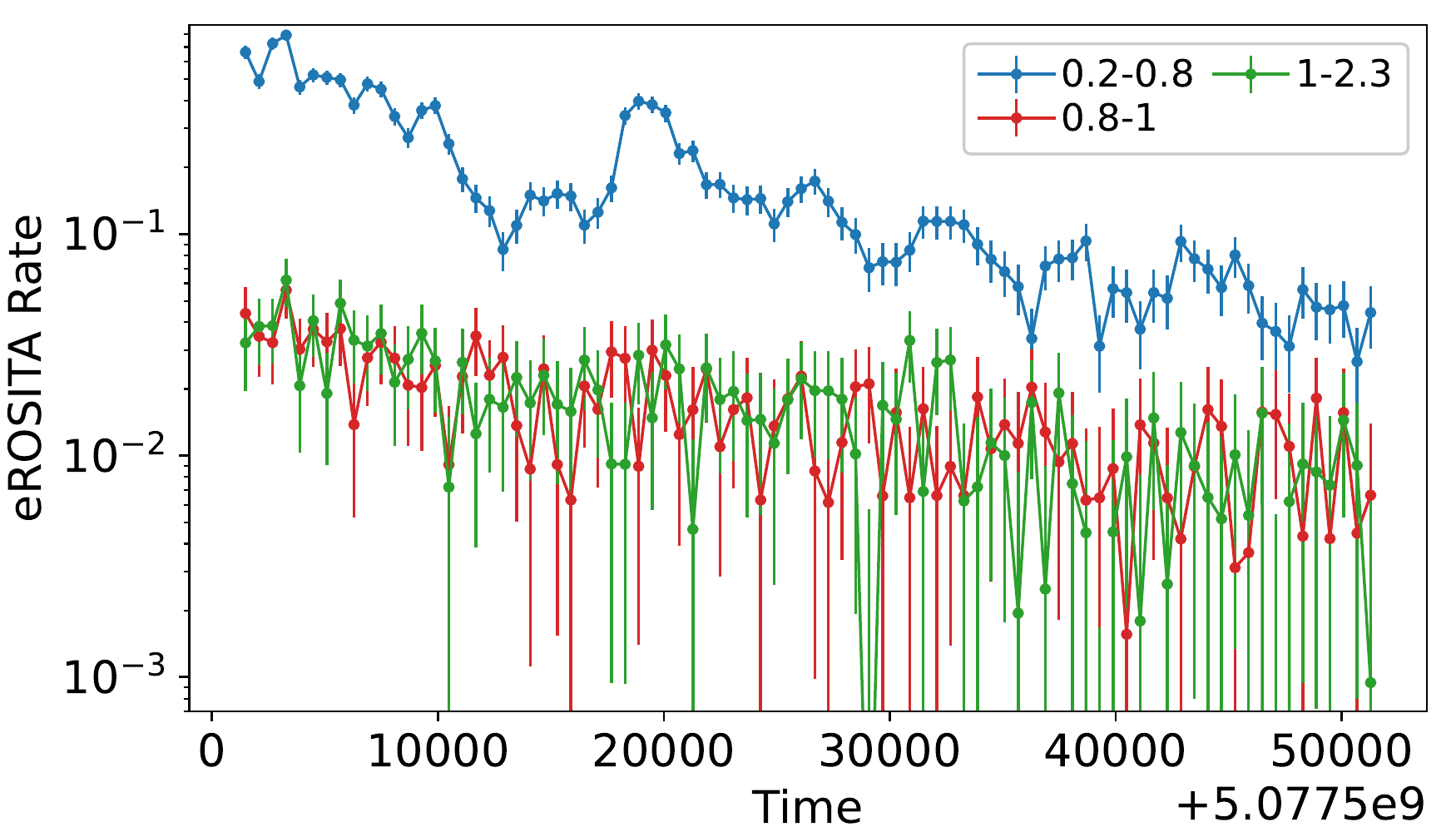}

    \caption{eROSITA light curve in three energy bands. The energy band up to only 0.8 keV is dominating the amplitude variability, followed by a sudden drop in variability above 0.8 keV.
}
    \label{fig:FigA3}
\end{figure}

We analysed the amplitude variability in each of the energy bands. In Fig.~\ref{fig:FigA3} we show the eROSITA light curves in three individual energy bands. The variability is dominant in the energy band up to 0.8 keV, with a sudden drop in variability  above 0.8 keV.

\section{eROSITA PV observation RGB image}

\begin{figure} 
        \includegraphics[width=\columnwidth]{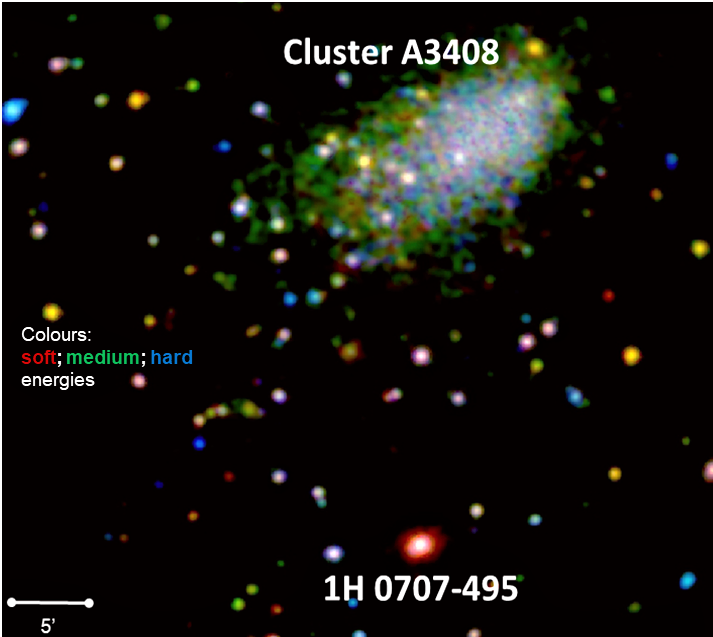}
    \caption{eROSITA RGB image of the field of view of the PV observations (
    0.2--0.7 (soft band, red), 0.7--1.5 (medium band, green), and 1.5--5 (hard band (blue)).
    Besides the primary target 1H~0707$-$495, the galaxy cluster A3408 has been detected for the first time with high spatial resolution. We also note the detection of a hard (blue) and probably obscured source population only detected above 1.5 keV.
}
    \label{fig:FigB1}
\end{figure}

In Fig.~\ref{fig:FigB1} we show the eROSITA image obtained during the PV phase observations. The objects are colour coded.
Thanks to the large field of view of eROSITA , the galaxy cluster A3408 was also serendipitously covered in addition to the super-soft source 1H~0707$-$495. This resulted in the best imaging information  for this cluster to date compared to ASCA observations \citep{2001Katayama}, revealing a very elongated morphology. The cluster is being studied in detail for a separate publication (Iljenkarevic et al., in prep.).

\end{document}